\documentclass[10pt,preprint2]{aastex}

\usepackage{framed,amsmath,ulem}

\def\kpch{h^{-1}{\rm kpc}}
\def\Mpch{h^{-1}{\rm Mpc}}

\def\Msunh{M_\odot\ h^{-1}}

\def\bk{\mathbf{k}}
\def\bx{\mathbf{x}}

\def\Np{N_{\rm p}}
\def\Nr{N_{\rm r}}
\def\Nrt{N_{\rm RT}}
\def\Ns{N_{\rm s}}

\def\Pop2{{\rm Pop II}}
\def\Pop3{{\rm Pop III}}

\usepackage{ng-defs}

\begin{document}

\title{Computer Simulations of Cosmic Reionization}
\author{Hy Trac\altaffilmark{1} and Nickolay Y.\ Gnedin\altaffilmark{2,3,4}}
\altaffiltext{1}{Harvard-Smithsonian Center for Astrophysics, Cambridge, MA 02138, USA} 
\altaffiltext{2}{Particle Astrophysics Center, Fermi National
  Accelerator Laboratory, Batavia, IL 60510, USA; gnedin@fnal.gov} 
\altaffiltext{3}{Department of Astronomy \& Astrophysics, The
  University of Chicago, Chicago, IL 60637 USA}  
\altaffiltext{4}{The Kavli Institute for Cosmological Physics, The
  University of Chicago, Chicago, IL 60637 USA}

\begin{abstract}
The cosmic reionization of hydrogen was the last major phase transition in the evolution of the universe, which drastically changed the ionization and thermal conditions in the cosmic gas. To the best of our knowledge today, this process was driven by the ultra-violet radiation from young, star-forming galaxies and from first quasars. We review the current observational constraints on cosmic reionization, as well as the dominant physical effects that control the ionization of intergalactic gas. We then focus on numerical modeling of this process with computer simulations. Over the past decade, significant progress has been made in solving the radiative transfer of ionizing photons from many sources through the highly inhomogeneous distribution of cosmic gas in the expanding universe. With modern simulations, we have finally converged on a general picture for the reionization process, but many unsolved problems still remain in this young and exciting field of numerical cosmology.
\end{abstract}


\section{Introduction}

Just as the 18th century explorers finished charting out most of the Globe, the 21st century explorers of the universe will most likely finish charting out the main areas of the cosmic evolutionary map. Already, the previous century has witnessed some major discoveries, and over the past two decades we have started converging on a ``\scm''. Two main ideas lay in the foundation of modern cosmology: the expansion of the universe and the formation of cosmic structure. The expansion of the universe was first discovered by Edwin Hubble in 1929 and it was recently found that the expansion is accelerating due to a mysterious dark energy \citep{cosmo:rfcc98,cosmo:pagk99}. This accelerated expansion continues to dilute the average density of matter in the universe,  giving space an increasingly empty appearance.

In contrast, structure formation involves gravitational contraction to higher densities. Tiny fluctuations in the density of matter at early times grew through gravitational instability to give rise to much larger cosmic structures at later times. The matter distribution, in which $80-85\%$ is dark matter and $15-20\%$ is cosmic gas, evolved to form a skeleton of high-density regions, called the ``large-scale structure" or ``cosmic web". Embedded in the highest density regions are the galaxies and clusters of galaxies that colour our picture of the cosmos. The large-scale distribution of galaxies in our cosmic neighbourhood has been cataloged by a number of surveys, culminating in the massive effort of the Sloan Digital Sky Survey (SDSS\footnote{http://www.sdss.org/}). While some properties of the galaxy distribution are well understood within the framework of the \scm, many questions remain about how galaxies form and evolve.

As modern cosmologists strive to document the important events in the cosmic history, they gain better understanding of how the two main ideas fundamentally shape the evolution of the universe. Currently, we have scoped out several periods of the cosmic evolutionary map, but some major epochs have not yet been well charted. In this review, we focus on the exciting exploration of the epoch of reionization, a frontier in modern cosmology.

In the cosmic chronology, the earliest observed information comes from the time when the universe was about 380,000 years old. At this stage, the cosmic plasma and radiation have cooled enough to allow electrons to combine with protons to form stable neutral hydrogen atoms. As a result, the primordial radiation mostly stopped scattering and since then has travelled largely unimpeded through the universe, to be detected by us as the cosmic microwave background. Measurements of the fluctuations or anisotropies in this radiation, first by the Cosmic Background Explorer (COBE\footnote{http://lambda.gsfc.nasa.gov/product/cobe/}) satellite, followed by many other experiments, and recently to unprecedented precision with the Wilkinson Microwave Anisotropy Probe (WMAP\footnote{http://lambda.gsfc.nasa.gov/product/wmap/}) satellite, have revealed much about the initial conditions of the early universe.

After the process of recombination, the next several tens of millions of years are referred to as the ``Dark Ages". During this period, the matter in the universe continued to evolve under the influence of gravity, eventually forming environments where early star-forming galaxies and rare quasars\footnote{Quasars are active nuclei of galaxies, powered by supermassive black holes. While they are located inside galaxies, they are often treated as a separate class of sources, as the spectrum of ionizing radiation that they produce differs substantially from the typical stellar spectra of ``normal'' galaxies.} were born. These luminous objects produced ultra-violet photons that are energetic enough to dissociate the electron from neutral hydrogen atoms. As the ionizing radiation escaped into the space between galaxies, called the ``intergalactic medium", the reionization of the cosmic gas begins.

The emergence of luminous sources marks the beginning of the epoch of reionization. A few hundred million years later, reionization is believed to be completed, for reasons which we discuss in more detail later. The cosmic reionization of hydrogen was the last major phase transition in the evolution of the universe. During that epoch, helium was also ionized, but typically only one of two electrons are dissociated from each atom. The full ionization of helium is believed to have occurred at a later time when quasars, with photons energetic enough to dissociate the second electron, become sufficiently abundant.

The study of reionization has emerged as a frontier topic in cosmology and is the focus of this review. We first discuss the current observational constraints in \S \ref{sec:observations} and then describe the dominant physical effects that control the ionization of the intergalactic gas in \S \ref{sec:physics}. In \S \ref{sec:methods} we present a qualitative description of the numerical methods for modeling reionization and in \S \ref{sec:simulations} we summarize results from modern computer simulations.

\section{Observational Constraints on Cosmic Reionization}
\label{sec:observations}

Cosmic reionization by early galaxies and quasars left some residual neutral hydrogen and an abundance of free electrons in the intergalactic medium. These tracers have been used to study the epoch of reionization. The residual neutral hydrogen is probed through Lyman alpha absorption in the spectra of high redshift quasars, while the free electrons are detected through Thomson scattering of the cosmic microwave background. We now review the current major observational constraints and later in \S \ref{sec:21cm} discuss how the neutral hydrogen can be observed directly through high-sensitivity radio observations.

\subsection{Lyman alpha absorption}

The first major constraint on cosmic reionization comes from observations of high-redshift quasars. The residual neutral hydrogen in the intergalactic medium can be quantified using Lyman alpha absorption\footnote{Since \lya\ absorption is resonant, the absorbed light eventually gets re-emitted; however, since it is re-emitted in a random direction away from the observing telescope, the scattering process appears as absorption to an observer. We, therefore, adopt the widely used (although inexact) term ``\lya\ absorption'' to describe this process.} features in the observed spectra. Most of the very distant quasars have been discovered in the SDSS, although a few have been found by other searches. We refer the reader to a comprehensive review by \citet{igm:fck06} for a recent discussion of observational searches for high-redshift quasars.

\begin{figure}[t]
\includegraphics[width=\hsize]{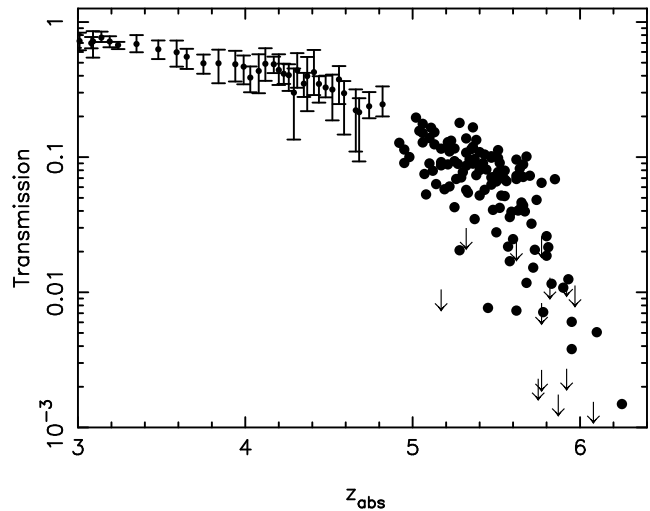}
\caption{\label{fig:lyaot} Measurements of the fraction of light from
  high redshift quasars, transmitted through the intergalactic medium;
  the rest of light is absorbed in the \lya\ line of neutral
  hydrogen. The rapid evolution at $z>5.5$ is commonly interpreted as
  the rapid evolution in the abundance of neutral hydrogen in the
  universe, possibly indicating the end of the reionization epoch at $z\sim6$
  \citep{igm:fck06}. Figure is courtesy of X.\ Fan.}
\end{figure}

The absorption measurements are summarized in Fig.\ \ref{fig:lyaot}, which we adopt from \citet{igm:fck06}. At intermediate redshifts ($2\lesssim z \lesssim 5$) the residual neutral hydrogen in the intergalactic medium causes absorption at the $\sim10-50\%$ level. The transmitted fraction gradually decreases at higher redshifts, but is still consistent with a highly ionized universe. However, the absorption increases much more rapidly between $z=5.5$ and $z=6$, and the few observations at $z>6$ suggest perhaps even more rapid change.

One commonly adopted interpretation of these observations is that the universe was much more neutral at the epochs probed by the higher redshift data. The transmitted fraction is observed to change by $\sim2-3$ orders of magnitude in only a small redshift interval $\Delta z\approx0.5$. The rapidity of the change, if extrapolated further, would suggest that the reionization epoch ended not long before $z=6$, perhaps somewhere between $z=6$ and $z=6.3$. This argument was presented shortly after the first high-redshift quasars were discovered by SDSS \citep{igm:bfws01}, and has since drawn additional support \citep[e.g.][]{igm:fnsw02, igm:wbfs03, ng:g04, igm:fsbw06, ng:gf06}.

However, the epoch of reionization is not directly probed by these observations. Since the \lya\ opacity of the fully neutral intergalactic medium at these redshifts is very large, of the order of $10^5$ or even higher, it only takes a small neutral fraction to have nearly complete absorption. The lowest measured transmitted fraction of $\sim10^{-3}$ at $z\approx6$ is consistent with a volume-averaged neutral fraction of only $10^{-4}$ to $10^{-3}$ \citep[e.g.][]{Lidz2006, Becker2007}. The evolution of the neutral fraction at higher redshifts and the exact timing of reionization remains, at present, unknown.

\subsection{Thomson electron scattering}

The second major constraint comes from observations of the cosmic microwave background radiation. As the universe evolves from the recombination epoch to the present time, the cosmic microwave background photons Thomson scatter with electrons dissociated during the reionization epoch. The scattering results in a small suppression of cosmic microwave background anisotropies on all scales and also generates additional polarization on large angular scales. A convenient quantity that parametrizes both effects is the total optical depth,  $\taut$, to Thomson scattering.

The WMAP satellite has detected the effects of Thomson scattering due to reionization and the first measurement of $\taut$ came after only one year of observations. The value reported, $\taut=0.17\pm0.04$ \citep{igm:ksbb03}, was unexpectedly large. Considering that the contribution to $\taut$ of the fully ionized intergalactic medium between $z=0$ and $z=6$ is only 0.04, a surprisingly large contribution, $0.13\pm0.04$, was left for the reionization epoch. Such a large value for $\taut$ would indicate an unusually prolonged or very early reionization epoch that is difficult to understand within the framework of the \scm.

However, recent measurements by WMAP after more years of observations are significantly lower with smaller uncertainties. The latest value, $\taut=0.087\pm0.017$, is based on 5 years of data, and is in good agreement with the models of reionization that are based on the modern state-of-the-art numerical simulations, as we discuss in \S \ref{sec:simulations}.

With a few more years of observations, WMAP will have slightly improved measurements of $\taut$. Furthermore, a much more precise measurement by the upcoming Planck Surveyor mission\footnote{{\tt http://www.rssd.esa.int/index.php?project=planck}} (to be launched in 2009) will allow the Thomson optical depth to be used to robustly constrain models of reionization.

\subsection{Lyman alpha emitters}

The third observational probe of reionization is provided by a class of high-redshift galaxies known as \lya\ emitters. Star-forming galaxies emit a significant fraction of their radiation at \lya\ wavelengths and these galaxies are recognized by the strong \lya\ emission lines. During the reionization epoch, the \lya\ emission from these galaxies is scattered by surrounding neutral hydrogen. The scattering changes the distribution of apparent brightness for the population of \lya\ emitters.

Changes in the abundance and distribution of \lya\ emitters can be used to detect the transition from the largely neutral to largely ionized intergalactic medium \citep[e.g][]{MalhotraRhoads2004, Kashikawa2006, Stark2007, McQuinnLAE2007}. However, recent research has shown that the interpretation of those measurements is extremely complex, and no consensus exists at present on how the observations of \lya\ emitters should be used to constrain reionization. Additional observations and better theoretical understanding of the effects of scattering on \lya\ emitters \citep[e.g.][]{Tasitsiomi2006, Dijkstra2007} are required to place more reliable constraints on the reionization epoch.

\section{Physics of Reionization}
\label{sec:physics}

In modeling reionization, the main important physical process is the transfer of ionizing radiation through the inhomogeneous distribution of cosmic gas. Since ionizing photons are not scattered, but only emitted and absorbed, and travel in straight lines between the acts of emission and absorption (with a minor and often negligible complication that they redshift as they travel in the expanding universe), the physics of reionization process is largely determined by the physical properties of sources and sinks for the ionizing radiation.

\subsection{Sources of ionizing radiation}
\label{sec:sources}

The nature of reionization sources remains poorly constrained. Two main types of astrophysical sources, massive stars in star-forming galaxies and quasars, are often considered likely to be the dominant ones. However, other more exotic possibilities, like decaying dark matter particles or evaporating primordial black holes, cannot be completely discounted yet. We will restrict our focus hereafter to galaxies and quasars as the only two choices of reionization sources since all simulations of reionization to date have only considered these cases. The reader must be cautioned, however, that the reality can be more complex than reionization modelers have been willing to accept so far.

\begin{figure}[t]
\includegraphics[width=\hsize]{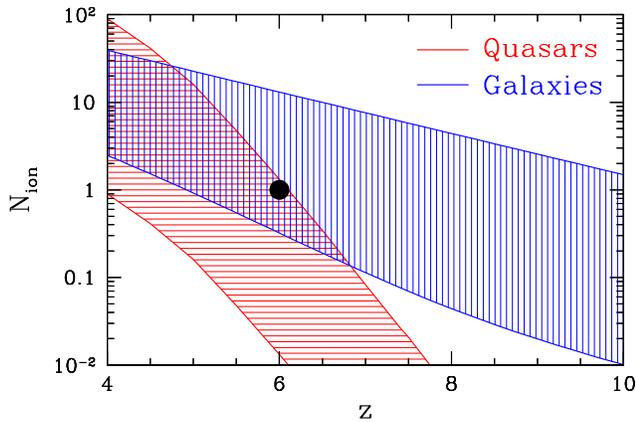}
\caption{\label{fig:ngb} Plausible ranges for the number of
  ionizations per hydrogen atom from quasars (red) and galaxies (blue) - the
  detailed explanations of adopted limits are given in the text. The
  large black dot is a necessary reionization conditions of one
  ionization per hydrogen atom by $z=6$.}
\end{figure}

The most basic condition on any type of source is that the total number of ionizations produced by these sources must at least equal the number of hydrogen atoms in the universe. Otherwise, it will not be possible for that kind of source to reionize the whole universe. This condition is actually only a minimum requirement. Since the cosmic gas is able to recombine, more than one ionizing photon per hydrogen atom is necessary to maintain the ionization. The required photon to atom ratio remains a major unknown - computing this quantity is an important task for the computer simulations that we describe below.

Fig.\ \ref{fig:ngb} shows a plausible constraint on the number of ionizations by galaxies and quasars. Since we lack a comprehensive theory of reionization, we show for each type of source a range of estimates that most likely bound the truth. The lowest quasar estimate is based on the direct extrapolation of the observed quasar luminosity function \citep{qlf:hrh07} to higher redshifts, assuming that each ionizing photon produces a single ionization. Quasars are believed to emit all of the ionizing radiation they produce and therefore, the total ionizing luminosity can be easily estimated by knowing their abundance. However, this extrapolation is likely to be an underestimate for two reasons. First, the spectrum of ionizing radiation from quasars is quite hard, with a large number of energetic photons and a mean photon energy of about one keV. When such an energetic photon ionizes a hydrogen atom, an energetic electron is produced. These energetic electrons can, in turn, ionize more atoms. Thus, a single ionizing photon can ionize more than one atom via the ``secondary ionizations'' process. While the exact number of secondary ionizations requires complex modeling, a rule of thumb for a crude estimate is that one ionization is produced for every 40 eV of the ionizing photon energy \citep{atom:ss85}.

Second, observations miss the low-luminosity objects that are fainter than the detection limits. In addition, the rarest bright ones may not be found within the finite surveyed volumes. Thus, the observations are likely to underestimate the total number of ionizing photons coming from all quasars. While the exact magnitude of this underestimate is difficult to compute, simple models indicate that it is unlikely to be more than a factor of 3 to 5 (M.\ Volonteri, private communication). The upper bound for the quasar range is therefore obtained by simply multiplying the lower bound by a factor of 100, and it should be considered an extremely conservative upper limit. Thus, Fig.\ \ref{fig:ngb} illustrates a conclusion that is well known since the pioneering work of \citet{igm:mhr99}; quasars alone are not powerful enough to reionize the whole universe by $z=6$, as required by the observations discussed in \S \ref{sec:observations}.

With star-forming galaxies the situation is more complex. On one hand, the luminosity functions of galaxies are measured with reasonable precision all the way to $z=6$, and some measurements exist up to $z\sim9$ \citep{hizgal:biff08}. The uncertainty due to incomplete observations is, therefore, not expected to be large
\citep{ng:g08}. On the other hand, galaxies do not emit all of the ionizing radiation they produce and the total ionizing luminosity can not be estimated by just knowing their abundance, unlike for quasars. Galaxies only leak a fraction of the radiation produced by their stars, with the rest being absorbed locally by gas clouds in the interstellar medium.

The fraction of ionization radiation that escapes from galaxies, $\fesc$, is a major uncertainty. The upper bound in Fig.\ \ref{fig:ngb} is computed from the observed  galaxy luminosity functions with the assumption that $\fesc=20\%$. This assumption may not be reasonable, however, since observations of bright galaxies at lower redshifts indicate that $\fesc\approx 2\%$. Thus, the upper bound for the stellar contribution to the ionizing budget should be, just as one for quasars, treated as an extremely conservative upper limit. The bottom bound for the stellar contribution follows the model of \citet{ng:gkc08}, which is based on the combination of observational data and numerical simulations of high redshift galaxies. If the simulations used by \citet{ng:gkc08} are not accurate enough, the lower bound may be an underestimate, although it is unlikely to be off by more than a factor of 3.

A simple conclusion can be drawn from Fig.\ \ref{fig:ngb}. Stars in galaxies are an important and most likely dominant source of ionizing radiation during the reionization epoch. However, if the escape fractions from galaxies are as low as is indicated by numerical simulations, then quasars can contribute as much as $\sim50\%$ of all ionizations of hydrogen.

\begin{figure}[t]
\includegraphics[width=\hsize]{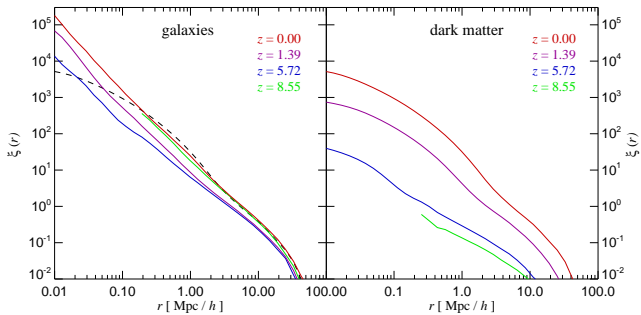}
\caption{\label{fig:bias} Correlation functions of galaxies (left) and
  dark matter (right) as a function of the comoving distance $r$ at
  four different epochs from the Millennium simulation of structure
  formation \citep{Springel2006}. Figure is reprinted with permission by V.\ Springel.} 
\end{figure}

To complicate things even further, galaxies and quasars are not distributed randomly in the universe, but populate a complex large-scale structure of clusters, groups, filaments, and voids\footnote{Qualitatively and even quantitatively, the distribution of matter in the universe is similar to the distribution of human population on the Earth. There are huge metropolitan areas with the population density orders of magnitude higher than the mean, large and small cities and towns strewn mostly along major highways, and small villages and hamlets in the most sparsely populated areas.}. The galaxies and quasars are clustered and aggregate together, such that the typical separation between nearest neighbors is smaller than what is expected if they are randomly distributed\footnote{If all the people on Earth were randomly distributed, the average distance to your nearest neighbor would be $160\dim{m}$, but most likely you will find someone much closer if you look around right now.}. The simplest measure of any clustered distribution is a ``correlation function", $\xi(r)$, which describes the fractional excess of neighbors at a given distance $r$ over a random distribution. In Fig.\ \ref{fig:bias} we show the galaxy correlation function at a range of redshifts as well as the correlation function for the dark matter at the same redshifts. These two functions are not the same as galaxies are always more clustered than the dark matter. The galaxies are said to be ``biased" and at high redshifts, the biasing can be quite large, as Fig.\ \ref{fig:bias} illustrates. In \S \ref{sec:reionizationprocess} we discuss the effects of the clustering of sources on the reionization process.

\subsection{Sinks of ionizing radiation}
\label{sec:sinks}

During the epoch of reionization, each neutral hydrogen atom serves as a sink for an ionizing photon, or at least a fraction of an ionizing photon if secondary ionizations are taken into account. If this was the only type of sink, the theory of reionization would have been completed by now.

Unfortunately for theorists, there exist other types of sink for ionizing photons. For example, the already ionized gas can recombine if its density is high enough. These recombinations will serve as an additional sink of ionizing photons, since each newly recombined atom would need to be ionized again to complete the reionization of the universe. It is generally believed that recombinations can consume a measurable but not the dominant fraction of all ionizing photons. They are an important sink for ionizing radiation at the
earliest stages of reionization, but become progressively less important as the universe expands and the gas density decreases.

Another type of sink for ionizing radiation, called ``Lyman limit systems'', exists both during and after the reionization epoch. Lyman limit systems are observed in the spectra of distant quasars as absorption systems with column densities of neutral hydrogen exceeding about $10^{17}\dim{cm}^{-2}$. At these column densities, the optical depth for ionizing radiation is $\tau_{LL}\gtrsim1$ at the Lyman limit wavelength $\lambda=912\mbox{\AA}$.

Usually, it is extremely difficult to make a connection between absorbing systems in quasar spectra and the physical objects in space that correspond to them. At present, astronomers know some properties of Lyman limit systems, for example, that they are mostly ionized and not highly neutral \citep{igm:p99}. However, we still lack the knowledge about what kind of physical objects they are, or how they correlate with galaxies and quasars.

For studies of reionization the most important factor is  the abundance of Lyman limit system as a function of redshift. After all, the only property of sinks of ionizing radiation is that they are indeed sinks, and that they absorb all ionizing photons that hit them, and it is not that important whether they are clouds of hydrogen gas or photon-eating space-dwelling monsters. For example, it generally makes little difference whether the Lyman limit
systems are highly clustered or distributed unformly in space, as long as they have the same abundance per unit redshift. The abundance mainly determines the probability for an ionizing photon to be absorbed, irrespectively of the true nature or spatial distribution of the absorbers.

The measurements of the abundance of the Lyman limit systems at $z<4$ has been presented by \citet{igm:smih94} some time ago, although more recent and accurate measurements from the SDSS should be forthcoming shortly. However, a more physically intuitive quantity is the ``mean free path" of ionizing radiation $\lambda_{\rm MFP}$, defined as\ the average distance an ionizing photon travels through space before being absorbed. The measured abundance of Lyman limit systems can be easily converted into the measurement of the mean free path of ionizing radiation. At $z\approx4$, $\lambda_{\rm MFP} \approx 90 h^{-1}\dim{Mpc}$
in comoving\footnote{Comoving distances are defined as real, physical distances scaled by $1+z$ to account for the expansion of the universe; they are convenient to use because comoving distances remain constant with time in an inertial reference frame. In addition, in order to scale out the dependence of cosmic distance on the insufficiently accurately known Hubble constant $H_0$, the comoving distances have been customarily expressed for many years in $h^{-1}\dim{Mpc}$, where $h=H_0/(100\dim{km/s}/\dim{Mpc})$.} units, with about 15\% uncertainty \citep{igm:m03}. 

At higher redshifts the constraints do not yet exist, but the observed evolution of the Lyman limit abundance is well described by a power-law in $(1+z)$, so it can be directly extrapolated to higher redshifts for a plausible, but uncertain, estimate of the mean free path from the Lyman limit systems at earlier times.

\subsection{General overview of reionization process}
\label{sec:reionizationprocess}

Armed with basic knowledge of sources and sinks for ionizing radiation,
we can try to understand general features of the reionization
process. While our goal is not to give a comprehensive review of the
theory of reionization, it is useful to understand the main process
and relevant spatial scales, as these are important for understanding
the limitations and the degree of scientific fidelity of simulations
that we describe below.

The story of reionization is not difficult to imagine. As sources of
radiation begin to produce energetic photons, they will start ionizing
regions around them, usually called ``\HII\ regions'' (after the
spectroscopic notation for ionizing hydrogen, \HII) or, more
colloquially, ``ionized bubbles''. The ionized bubbles keep expanding
until... well, until that expansion stops.

There could be two physical reasons why the expansion stops: either
because all the ionized bubbles overlap, and all of the intergalactic
medium gets reionized, or because the Lyman limit systems inside the
bubbles absorb all ionizing photons available for ionizing gas outside
a bubble. Since the bubbles originate around sources, the choice
between the two scenarios is determined by the relationship between
the mean free path for ionizing radiation and the average distance
between ionizing sources. If the source separation is much smaller
than the mean free path (``abundant sources'' scenario), then most of
ionizing bubbles overlap before the absorptions by Limit limit systems
becomes significant, the overlap of bubbles happens fast, and results
in a quick reionization of the universe \citep[i.e.][]{ng:g00a}.

If the opposite is true, and the sources are so strongly clustered
that their correlation length is much larger than the mean free path
(``rare sources'' scenario), then ionized bubbles reach their maximum
sizes (comparable to a few times the mean free path) before they can
overlap, and reionization process stalls
\citep{igm:mhr00,igm:fm09}. The whole universe can then be reionized
only when either the abundance of Lyman limit systems decreases (due
to the cosmic expansion and the associated density decrease or because
they get ionized more) or sources become more numerous or less
strongly clustered. In other words, reionization does not actually get completed in a pure 
``rare source'' scenario. Only when sources become sufficiently
abundant, with at least one source (or a group of clustered sources)
per every sphere with radius of a few $\lambda_{\rm MFP}$, can
reionization of the whole universe end.

\begin{figure}[t]
\includegraphics[width=\hsize]{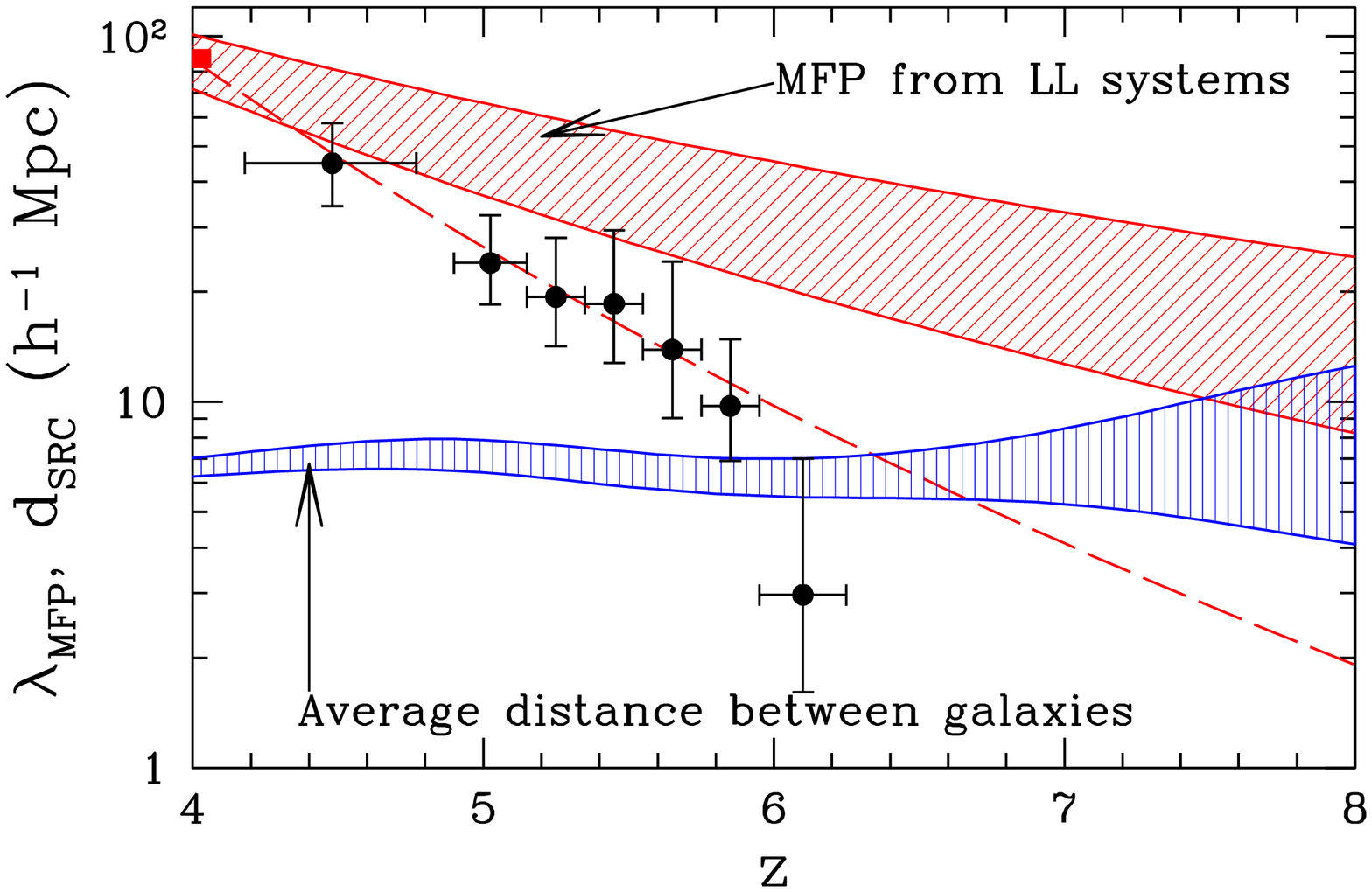}
\caption{\label{fig:mfp} The estimate of the mean free path from the
  Lyman limit systems as an extrapolation of the observed evolution of
  the Lyman limit abundance \citep[][; red band]{igm:smih94}. The blue
  band shows the mean galaxy separation as a function of redshift
  \citep{hizgal:biff07,hizgal:biff08}. Black dots with error-bars are
  observational 
  estimates of the mean free path from the spectra of SDSS quasars
  \citep{igm:fsbw06}. The dashed red line shows an extrapolation of
  the mean free path from the Lyman limit system that matches the SDSS
  estimates.}
\end{figure}

In principle, these two scenarios can be distinguished by
comparing the abundance of Lyman limit system (i.e.\ the mean free
path) and the abundance of sources during reionization. In practice,
the situation is rather complex. The current observational constraints
are summarized in Fig.\ \ref{fig:mfp}. There we show the observed mean
free path from Lyman limit systems (red square at $z=4$) and its
extrapolation to higher redshifts (red hatched band) using the
observed rate of evolution of Lyman limit systems at lower redshifts
\citep{igm:smih94}, the estimate of the mean separation between
galaxies from recent surveys of Lyman Break galaxies
\citep{hizgal:biff07,hizgal:biff08}, as well as observational
estimates of the mean free path for ionizing radiation from the
spectra of SDSS quasars \citep[][black points with
  error-bars]{igm:fsbw06}\footnote{The red hatched area is computed
  assuming $dN/dz = (3.3\pm0.5)\left((1+z)/5\right)^{2.55\pm0.65}$;
  the blue hatched area is obtained from \citet{hizgal:biff08} fits as
  $\phi_*^{-1/3}$; the dashed red line assumed the Lyman limit system
  evolution in the form $dN/dz = 3.3\left((1+z)/5\right)^6$.}.

Taken at face value, the existing constraints seem to suggest that
the mean free path is larger than the mean galaxy separation, and that
the ``abundant sources'' scenario is realized. For example, the fact
that the estimate of the mean free path from the spectra of SDSS
quasars drops rapidly at $z\approx6$ has been used to argue that the
fast overlap of ionized bubbles occurred at that redshift \citep{ng:gf06}.

The situation, however, is likely to be more complex. First of all,
both the mean free path $\lambda_{\rm MFP}$ and the mean galaxy
separation $d_{\rm SRC}$ should be taken as characteristic scales, not
as exact distances. Since the mean galaxy separation is only a factor
of few below the extrapolated mean free path, it is not really in the
regime where it is ``much smaller'' than $\lambda_{\rm MFP}$. 

In addition, galaxies are highly clustered, as Fig.\ \ref{fig:bias}
illustrates. In a highly clustered distribution, the mean distance
between galaxies is a poor indicator of actual spacing between
sources: some galaxies will be so close to each other that they
effectively form a single, more powerful source, while clusters of 
sources can be spaced by much more than $d_{\rm SRC}$. For example, in
the present day universe the average distance between galaxies is about
$4-5h^{-1}\dim{Mpc}$; never-the-less, voids in the large-scale
distribution of galaxies reach sizes of several tens of Mpc and
filamentary super-clusters approach hundreds of Mpc in
length.

It is also possible that our extrapolation of the mean free path from
the Lyman limit systems is incorrect. For example, if the abundance of
the Lyman limit systems is larger at higher redshifts, the mean free
path would be smaller. As an illustration, the dashed red line in
Fig.\ \ref{fig:mfp} shows the extrapolation of the Lyman limit system
abundance that matches well the estimate of the mean free path from
the SDSS quasars. In
that case the mean free path gets smaller than 
the galaxy correlation length at $z\approx6.5$. 

Thus, we must conclude that the existing knowledge of the reionization
epoch does not clearly prefer either abundant or rare sources
scenario; the reality is, most likely, to be somewhere in 
between the two extremes. That makes the theorist's task of creating
simple, analytical theories of reionization so much harder, and points
toward detailed numerical simulations of reionization as the
ultimate method of choice.

\section{Numerical Methods}
\label{sec:methods}

Computer simulations provide a powerful and versatile tool for solving the fundamental physics of gravitation, gas dynamics, and radiative transfer, which are important for modelling reionization. Many specialized algorithms have been developed and applied to simulating the complex interactions between dark matter, gas, stars, and
radiation. Numerical modeling has also benefitted from the rapid development in supercomputing technologies. Computing capabilities, such as processor speed and memory capacity, have been growing according to Moore's Law, doubling approximately every two years.  Together, these advancements enable increasingly larger and more realistic simulations to be run.

With modern cosmological simulations, we can now study how the small, initial perturbations in the dark matter and cosmic gas distributions undergone nonlinear gravitational collapse to form the large-scale structure of the universe \citep[see][for reviews]{Bertschinger1998, Springel2006}. The large-scale structure, characterized by filaments and halos, sets up cosmic environments where the early stars and galaxies form. While astrophysical processes such as star formation still can not be simulated from first principles, simulations do allow a more straightforward implementation of intricate prescriptions, motivated by theory and calibrated against available observations.

Over the past two decades, there have been considerable interest in cosmological simulations with radiative transfer. Earlier work focussed on solving for the radiation field around a single point source, which is applicable to the first stars or first galaxies. In the last few years, the attention has shifted towards understanding how the larger distribution of early stars and galaxies photo-ionized and photo-heated the intergalactic medium during the epoch of reionization. Radiative transfer in cosmological simulations is still in its infancy compared to dark matter and gas dynamics, but it is rapidly gaining strength and scope. In the following sections, we review the computational methodology and requirements for simulating cosmic reionization.

\subsection{N-body and Hydrodynamic Algorithms}

Currently, there are two classes of cosmological simulations. One class, designated ``N-body" simulations, models the evolution of all matter in the universe as a collisionless fluid influenced only by gravitational dynamics. The other class, called ``N-body + hydro" or just ``hydro" simulations, models both the collisionless dynamics of dark matter and the collisional dynamics of cosmic gas.

N-body algorithms are normally used to evolve the dark matter distribution, which is discretized into $N$ number of particles of fixed mass, each with known position and velocity. Newton's equations of motion can be solved in discrete time steps when given the acceleration. The exact force on a given particle can be calculated by doing a direct summation of the pair-wise forces exerted by all other particles. However, this is prohibitively expensive for cosmological simulations because the number of calculations scales as ${\cal O}(N^2)$ where $N$ can be large, currently up to 70 billion particles \citep{Teyssier2009, Kim2008}. Fortunately, several successful techniques have now been implemented to solve Poisson's equation for gravity with reasonable ${\cal O}(N\log N)$ scaling \citep[see][for a review]{Bertschinger1998}.

Hydrodynamic algorithms solve the fluid equations for the cosmic gas,
using grid-based (``Eulerian'') or particle-based (``Lagrangian'') techniques. In the Eulerian approach, the conservation equations for gas mass, momentum, and total energy are solved on a structured or unstructured grid of cells. The mass in a cell can have practically any value and hence grid-based algorithms are said to have high dynamic range in mass. Eulerian codes using uniform grids are usually more suited to simulating the intergalactic medium, which contains both dense and underdense gas. However, uniform grids have limited spatial dynamic range. Adaptive mesh refinement codes additionally have the capability to resolve small-scale structure like gas in halos.

In the Lagrangian approach, fluid elements are represented by particles of fixed mass and smooth particle hydrodynamics \citep[SPH;][]{Lucy1977, GingoldMonaghan1977} solvers are used to follow their trajectories. For each collisional particle, the dynamical forms of the fluid equations for density, velocity, and temperature are solved while smoothing over a small number of neighboring particles. SPH is especially suited for simulating
small-scale structure because the Lagrangian flow naturally allows for high spatial dynamic range in high density regions. On the other hand, underdense regions in the intergalactic medium are resolved with fewer particles.

Hydrodynamic algorithms can be run simultaneously with N-body algorithms to couple the gas and dark matter through their mutual gravitational attraction. Cosmological simulations of reionization also solve for atomic processes such as ionization, recombination, cooling, and heating. The photo-ionization and photo-heating of the gas are straightforward to calculate given the inhomogeneous radiation field.

\subsection{Radiative Transfer Algorithms}
\label{sec:rt}

\begin{figure*}[t]
\begin{center}
\includegraphics[height=0.25\hsize]{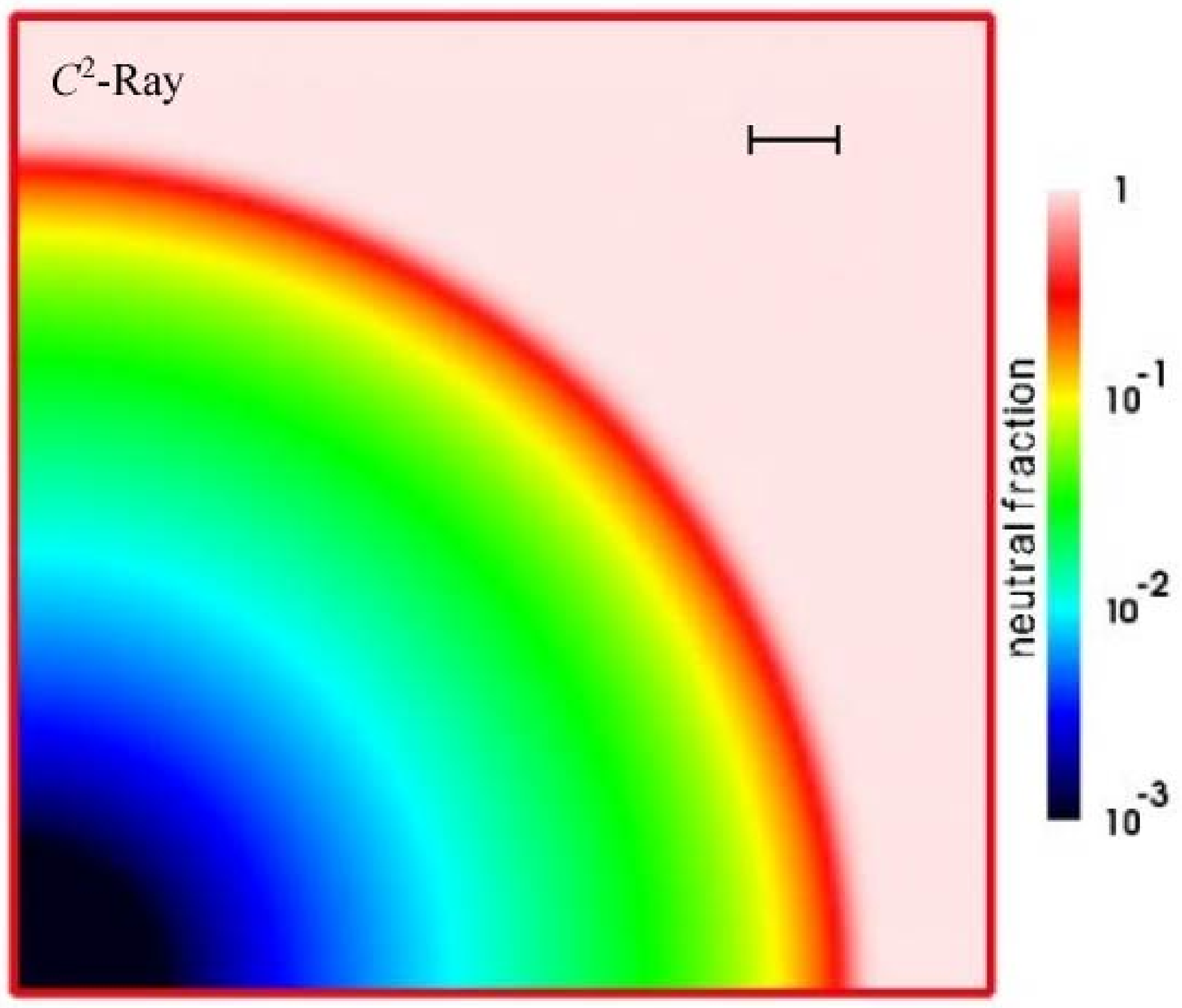}
\includegraphics[height=0.25\hsize]{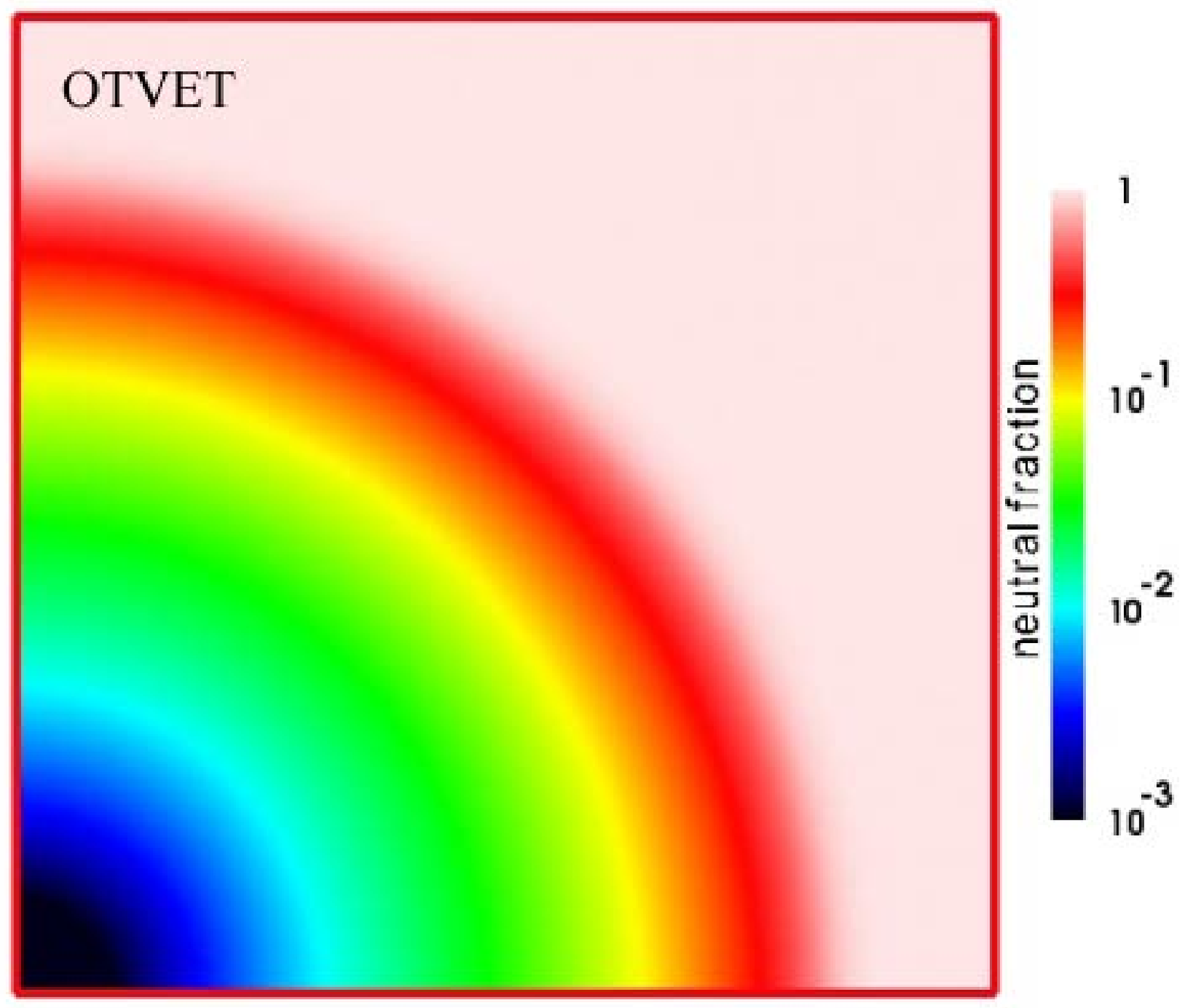}
\includegraphics[height=0.25\hsize]{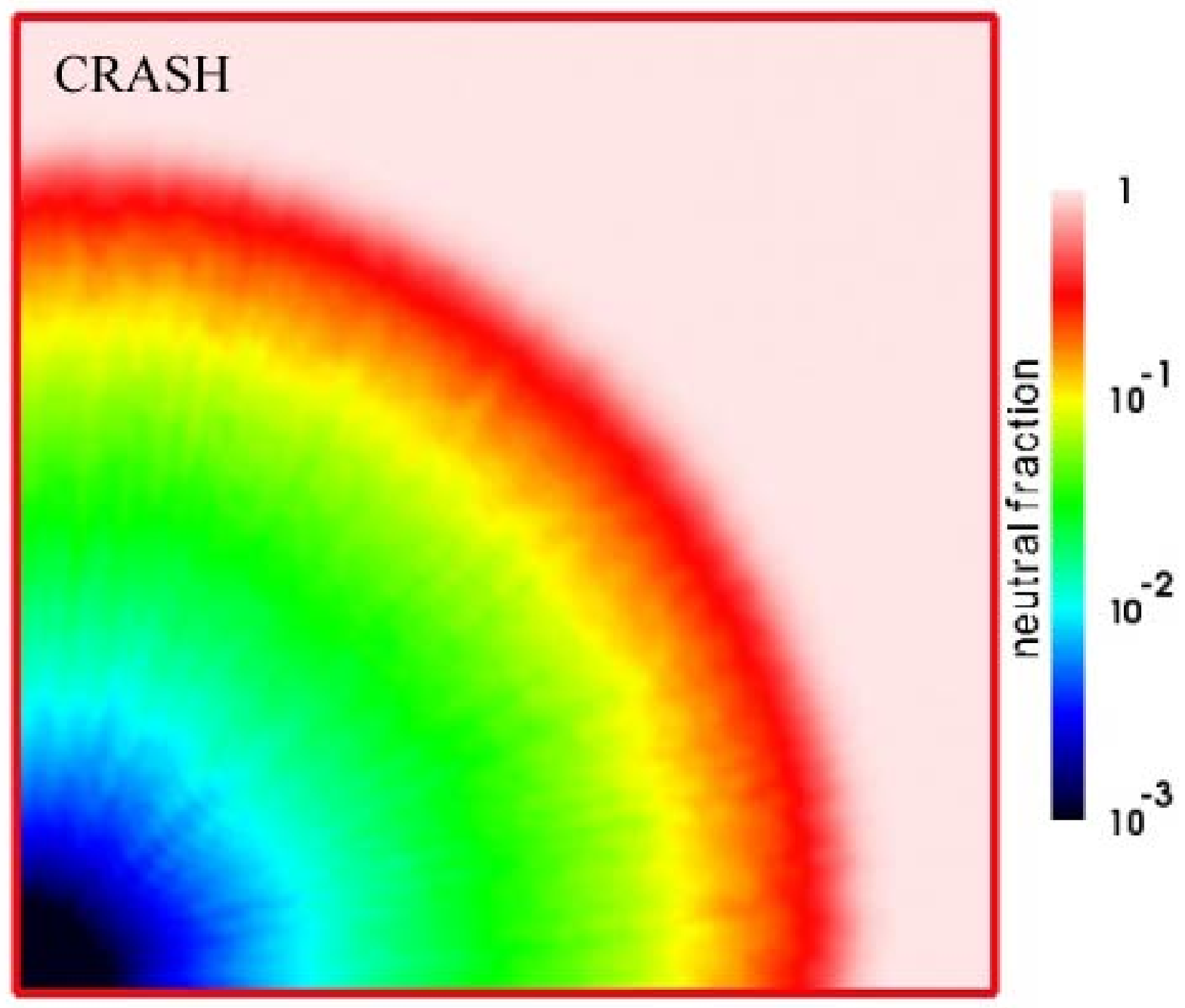}
\end{center}
\caption{\label{fig:rtcodes} A comparison of different numerical approaches for modeling radiative transfer by  \citet{IlievRT2006}. Each panel shows an octant of a spherical ionization bubble blown up by a single source, with the approach/code used in the simulation labeled above. For this classical test, the ray-tracing algorithm (C$^2$-RAY) provides the most accurate solution; the moment method (OTVET) is usually somewhat diffusive, while the Monte-Carlo approach (CRASH) introduces random fluctuations that break the perfect symmetry of the problem. Figure is courtesy of I.\ Iliev.} 
\end{figure*}

Radiative transfer algorithms solve the evolution of the radiation field, taking into account emission, absorption, and scattering processes. In general, the evolution is described by a differential equation for the specific intensity, which is a function of seven variables: 3D position, 2D angular coordinates, time, and frequency. Because of the high dimensionality of the problem, direct numerical solutions are computationally difficult and expensive.

For a discrete number $\Nrt$ of radiative transfer resolution elements, a direct solution requires ${\cal O}(\Nrt^{5/3})$ operations per frequency bin per time step. With this costly scaling, only low resolution simulations can be run if a brute-force solution is attempted. In order to be feasible for use in cosmological simulations, radiative transfer algorithms should scale close to linearly with the number of resolution elements, just like good N-body and hydrodynamic algorithms. However, satisfying this criterion requires some level of physical approximations and computational optimizations.

Existing algorithms can be broadly divided into three categories: moments, Monte Carlo, and ray-tracing methods. Recently, \citet{IlievRT2006} conducted a comparison of 11 cosmological radiative transfer codes using 5 simple test problems and found good general agreement between different algorithms. Figure \ref{fig:rtcodes} shows some results from a classical test of an ionized bubble expanding from a single source into a gas of uniform density and temperature. Below, we provide a qualitative description of the three main methods for modeling cosmological radiative transfer.

\subsubsection{Moments methods}

The radiative transfer equation for the specific intensity can be simplified by considering moments of the radiation field. It can be reduced to a simpler system of conservation equations for the photon energy density and flux. This is analagous to replacing the Boltzmann equation for the fluid distribution function with the Euler conservation equations for gas mass, momentum, and total energy. Furthermore, similar to the Euler equations having source terms on the right-hand side, for example coming from gravity, the radiative transfer moments equations have three important terms, two of which come from the emission and absorption. The third term, called the Eddington tensor, is related to the radiation pressure and is necessary to close the system of partial differential equations. 

Radiative transfer moments algorithms are naturally coupled to Eulerian hydrodynamic codes. Since the radiative transfer moments equations resemble the Euler hydro equations in form, they can be solved using the same, well-established techniques. Algorithms generally scale as ${\cal O}(\Nrt \log \Nrt)$, independently of the number of sources. The source function can easily incorporate both point-source and diffuse radiation. This advantageous feature is not generally shared by other methods. 

\citet{GnedinAbel2001} were the first to develop an algorithm specifically for reionization and proposed the Optically Thin Variable Eddington Tensor (OTVET) approximation. Because computing the exact Eddington tensor is a highly nontrivial task, the OTVET approximation computes the tensor under the assumption that the absorptions are negligible, but then used this ``optically thin'' tensor in the full, ``optically thick'' equation for the radiation energy density and flux. That ensures the conservation of photon number and flux, but causes errors in the direction in which the radiation flux is advected. One positive is that these errors remain under control at all
times, and the accuracy of the OTVET approximation can always be estimated; if this accuracy is found to be inadequate, a different, more precise method must be used instead. Fig.\ \ref{fig:rtcodes} demonstrates that, overall, the OTVET approximation agrees well with other techniques, but, just like Monte Carlo methods, is somewhat more diffusive than direct ray-tracing algorithms.

Several alternative schemes for computing the Eddington tensor have been developed recently \citep[e.g.][]{AubertTeyssier2008, Finlator2009, PetkovaSpringel2008}. These various algorithms mainly differ in how the Eddington tensor is computed, and what assumptions are made in that computation.

\subsubsection{Monte Carlo methods}

In Monte Carlo methods, the radiative transfer equation is solved using a probabilistic technique where the values of variables are randomly sampled from known probability distribution functions. The radiation field is discretized using photon packets and a number of packets is emitted from each source. For each packet, initial conditions such as the emission location, propagation direction, and frequency are determined by randomly sampling from the appropriate probability distribution. As a packet is transported away along a radial path, it will intersect gas elements and for each crossing, a fraction of the photons is consumed based on the probability for absorption.

For a single source, the number $\Np$ of packets emitted should be comparable to the number $\Nrt$ of radiative transfer resolution elements in order to have information about the radiation field everywhere. For a general problem with $\Ns$ sources, the total number of packets emitted is then $\Np\Ns$. In reionization simulations, $\Ns$ can be very large, scaling linearly with $\Nrt$, and thus making the overall scaling be ${\cal O}(\Nrt^2)$. In practice, optimizations are introduced such that the total number of packets is proportional to the number of resolution elements and only ${\cal O}(\Nrt)$ operations are done. This is usually achieved by reducing $\Np$ and comes at the expense of degrading the angular resolution around sources. Convergence tests, where $\Np$ is varied for example, must then be conducted to ensure that statistical fluctuations do not significantly affect results. With good optimization, Monte Carlo methods can be efficient and reliable.

CRASH \citep{Ciardi2001, Maselli2003} was the first Monte Carlo code written for reionization. The latest version operates on static, grid-based density fields and solves for the time evolution of H and He ionization fractions and gas temperatures. When compared with other cosmological radiative transfer codes, it showed good overall agreement on test problems, but tended to have thicker ionization fronts \citep{IlievRT2006}.  In Fig.\ \ref{fig:rtcodes} the broadening of the ionization front comes from the probabilistic nature where there is always a finite chance that a photon packet stops before or travels beyond the actual ionization front. More recent implementations \citep[e.g.][]{Semelin2007, Altay2008} have been designed to couple to SPH, allowing better spatial resolution in high density regions.

\subsubsection{Ray-tracing methods}

Ray-tracing is the most popular of the three categories and there is a rich diversity of approaches. We first review the basic techniques and then discuss adaptive improvements. For the basic case, we consider ray-tracing from a single source through a radiative transfer grid \citep[e.g.][]{Abel1999}. A fixed number of rays is generated with an isotropic distribution of propagation directions. Walking downstream away from the source, each ray is cast into segments as it traverses the radiative transfer grid, basically one segment for every cell intersected. Photons are consumed in each segment based on the local optical depth for photo-ionization, which depends on the length of the segment and the density of absorbers in the cell. Each ray is calculated independently, and gets terminated when its photon count goes to zero.

In a multi-source problem, the situation is more complicated because rays cannot be calculated independently in general. For a causal solution, all rays must be synchronized and traced in discrete steps, by one cell length at a time over a time step equal to the time it takes light to cross a cell. This gives the most accurate results, but is more costly as many steps are needed because the light-crossing time can be quite small compared to the duration of reionization. However, practical but approximate solutions do exist where sources can be processed independently, allowing for much longer time steps \citep[e.g.][]{Sokasian2001, Mellema2006}. 

Ray-tracing is computationally expensive and also runs into the ${\cal O}(\Nrt^2)$ scaling problem we discussed earlier for Monte Carlo methods. In the brute-force approach, the number $\Nr$ of rays cast per source must be equal to or greater than the number $\Nrt$ of radiative transfer resolution elements. This is necessary to have rays intersect cells far from the source, but it results in much more rays than necessary nearer to the source. Fortunately, there are adaptive improvements to make algorithms run faster and scale better. 

\citet{AbelWandelt2002} suggested an adaptive ray-splitting technique where the angular resolution continually increases farther away from sources. For a given source, a small number of parent rays are cast and travel a short distance before they split into 4 daughter rays. Successive generations of splitting continue downstream. Furthermore, adaptive ray-tracing algorithms can be made to scale more linearly with the number of resolution elements by grouping neighboring sources \citep{RazoumovCardall2005}, limiting the splitting of rays \citep{McQuinnHII2007}, or merging near-parallel rays \citep{TracCen2007}.

So far we have only discussed ray-tracing techniques for grid-based hydrodynamic codes. Several ray-tracing algorithms have also been developed for cosmological SPH simulations \citep[e.g.][]{Alvarez2006, Susa2006, PawlikSchaye2008}. The ray casting techniques in SPH are different because of the irregular distribution of the particles. One common approach is to emit a fix number of rays from each source and collect them using an unstructured grid based on the particle distribution. High spatial dynamic range is achieved for both the particles and the rays in high density regions.

C$^2$-RAY \citep{Mellema2006} is one example of a grid-based ray-tracing code that has been used for simulating reionization. This photon-conserving code traces rays away from every source to every radiative transfer cell. While each source is processed independently and in random order, an iterative procedure is used to converge to a causal and accurate solution. The current version scales linearly with the number of sources, which is costly when $\Ns$ scales with $\Nrt$. In Fig.\ \ref{fig:rtcodes}, the ionized bubble is in good agreement with the analytical solution and with results from other ray-tracing codes \citep{IlievRT2006}.

\subsection{Physical Scales and Computational Requirements}

\begin{figure}[t]
\includegraphics[width=\hsize]{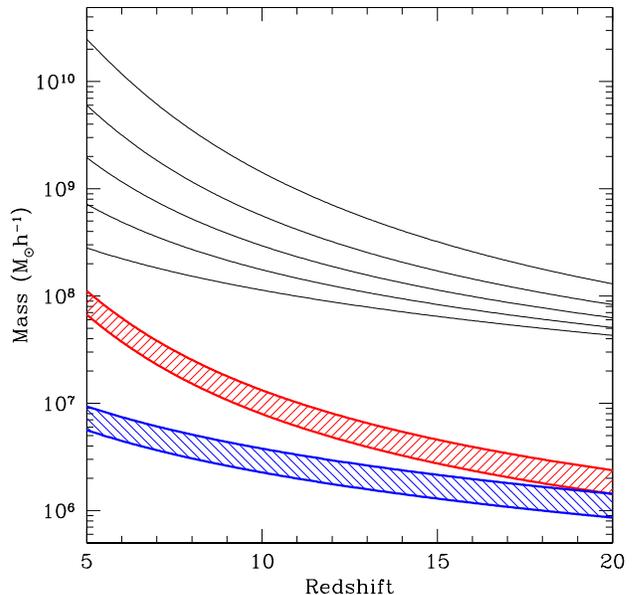}
\caption{\label{fig:massres} Mass range of dark matter halos that can host galaxies. The curve $M(z,f)$ gives the mass range $M\geq M(z,f)$ that accounts for the fraction $f$ of the expected total luminosity at redshift $z$. The five solid curves, from top to bottom, correspond to $f=$ 0.2, 0.4, 0.6, 0.8, and 1.0. If 30 to 50 particles are required to identify a halo in an N-body simulation, then the red and blue bands specify the particle mass resolutions that are needed to account for 50\% and 100\% the total luminosity, respectively.} 
\end{figure}

Computer simulations of reionization require a large dynamic range in both length and mass scales. Even with modern supercomputers, only a portion of the entire range of scales can be probed by any single simulation. High resolution is required to resolve small-scale structure such as radiation sources and sinks, the two key factors regulating reionization. Large simulation volumes are necessary to have a fair representation of the distribution of galaxies and ionized bubbles, and to find rare quasars. 

For studying the large-scale properties of reionization, the minimum simulation box size should be approximately 100 $\Mpch$ on a side for two major reasons. First, as we discussed in \S \ref{sec:physics}, the mean free path for ionizing photons is expected to be several tens of Mpc at $z\sim6-10$. Thus, the simulation box size needs to be many times larger in order to have a fair sampling of the distribution of ionized bubbles. Second, \citet{BarkanaLoeb2004} showed that a box size of about $100\ \Mpch$ is necessary to have a fair sample of dark matter halos where sources reside. Within this volume, the statistical fluctuations in the abundance of galaxies are small enough that is representative of the universe on average. Therefore, this volume is reionized no earlier or later than the universe on average by any significant amount.

Quasars are much rarer than galaxies, especially at early times. The quasars observed at $z\sim6$ in the SDSS have a very low number density, roughly one for every $(1000\ \Mpch)^3$ of space \citep[e.g.][]{igm:fck06}. Large-scale cosmological N-body simulations are able to find massive dark matter halos with mass $\sim10^{13}\dim{M}_\odot$, which host these quasars \citep{Li2007}. If the upper limit on the required box size is taken to be $1000\ \Mpch$, then the total mass within this volume, $\sim10^{20}\dim{M}_\odot$ or about $10^8$ times the mass of all matter in our galaxy, sets the corresponding mass limit.

On small-scales, the situation is rather complicated. Fig.\ \ref{fig:massres} shows the mass range of dark matter halos that is expected to host galaxies. For the redshift range of interest, halos with mass scale $\sim10^8\ \Msunh$ and comoving length scale $\sim10\ \kpch$ are thought to be able to form stars more efficiently. Within these halos, the gas reaches an important temperature scale $\sim10^4$ \dim{K} at which it can cool rapidly through atomic transitions. The dissipation of energy enables further gravitational collapse to much higher densities, allowing the formation of giant molecular clouds and in these the formation of stars. This sets the upper limit for what the minimum spatial and mass resolutions should be in order to locate where galaxies reside. The resolution has to be much better in order to resolve the gas structure within these halos. For an extreme example, we have to resolve down to tens of proper parsecs to directly form giant molecular clouds of order a thousand solar masses in gas.

It is difficult to set resolution limits for radiation sinks because we do not fully understand what the absorbing systems are. \citet{ng:kg07} have found that Lyman limit systems at $z=4$ are generally found near galaxies. They are typically a \dim{kpc} in proper size and have densities roughly a thousand times larger than the universal average. While this gives us a sense of the required scales, the situation at higher redshifts could be different. For example, consider halos that were not massive enough to form stars efficiently. These mini-halos were very abundant compared to the source halos and their reservoir of absorbing gas could make them significant radiation sinks \citep[e.g.][]{Haiman2001, Shapiro2004}.

Having some idea of the physical scales, we consider the demand in dynamic range and computational resources. The extreme scenario, simulating the formation of giant molecular clouds within large cosmic volumes where rare quasars can be found, would require over 7 orders of magnitude in length and 16 orders of magnitude in mass. While adaptive simulations can achieve this dynamic range for any given galaxy, it is still far out of reach to accomplish this for every galaxy within a very large cosmic volume. 

For a more practical case, we consider simulations that capture the formation of the large-scale structure and the reionization of the intergalactic medium within a simulation box size of $100\ \Mpch$. Given the distribution of halos that host galaxies, radiation sources and sinks can then be modeled approximately by populating these halos with assumed distributions for properties such as source luminosities and sink opacities. For example, the source luminosity can be assumed to be proportional to the halo mass. In N-body simulations, approximately 30 to 50 particles are needed to properly identify a dark matter halo and measure its mass. To form halos with $M = 10^8\ \Msunh$ then requires approximately 20 to 35 billion particles in total within the specified volume. Only recently has this been achieved by high-resolution N-body simulations utilizing 512 to 2048 processors, 2 to 4 trillion bytes (terabytes) of memory, and 100 to 200 thousand cpu-hours on modern supercomputers \citep{Shin2008, IlievTG2008, TracCL2008}.

Ideally, an equal number of hydrodynamic elements, either Eulerian grid cells or SPH particles, would be used to represent the cosmic gas. Hydro simulations have not yet been run at this scale because they generally require at least twice as much computer memory and take over ten times longer to run compared to N-body simulations. The computational costs will also increase substantially with the addition of radiative transfer. This major milestone will be achievable within the next few years using supercomputers with of order 10,000 processors and tens of terabytes of memory.

\section{Computer Simulations}
\label{sec:simulations}

\begin{figure*}[t]
\includegraphics[width=0.33\hsize]{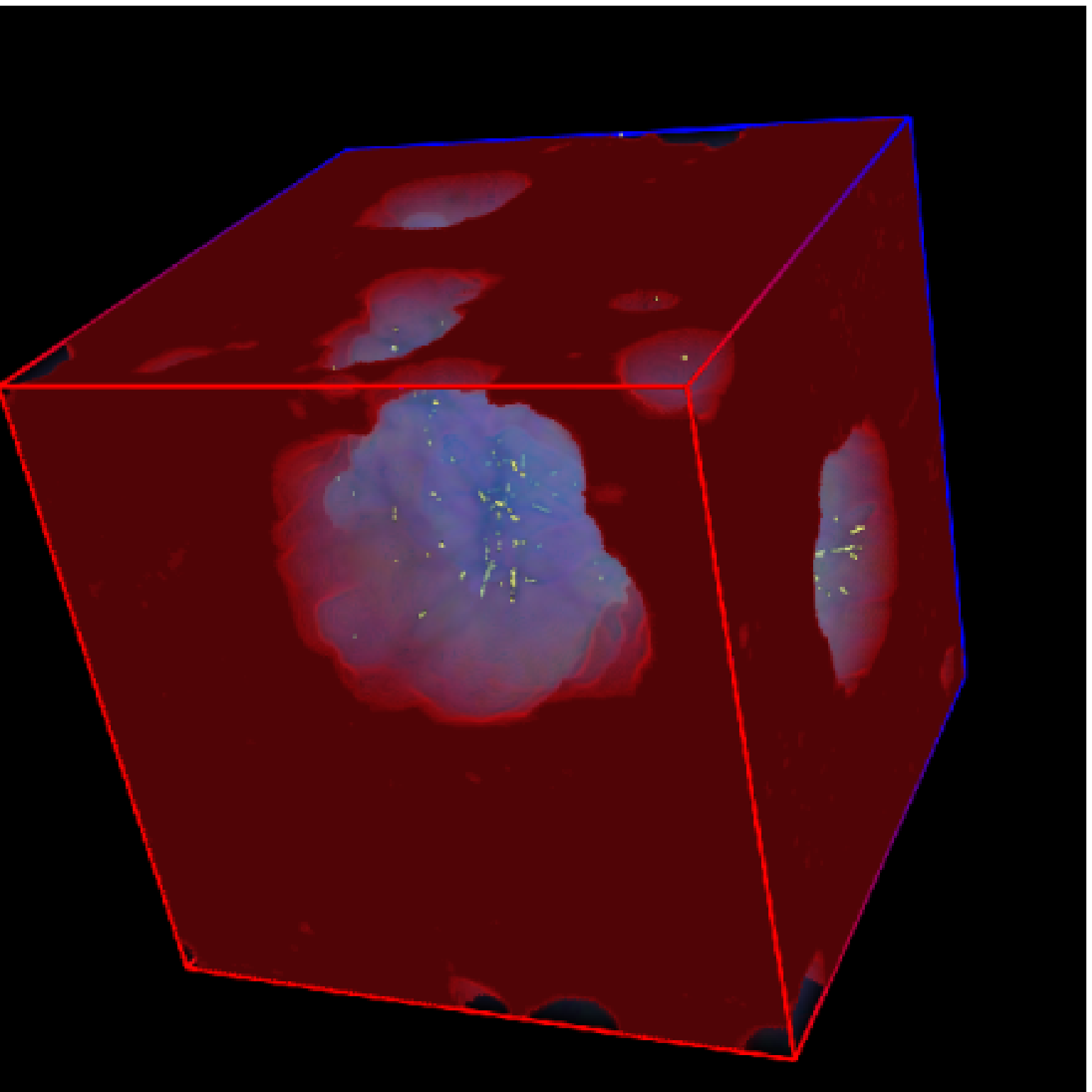}
\includegraphics[width=0.33\hsize]{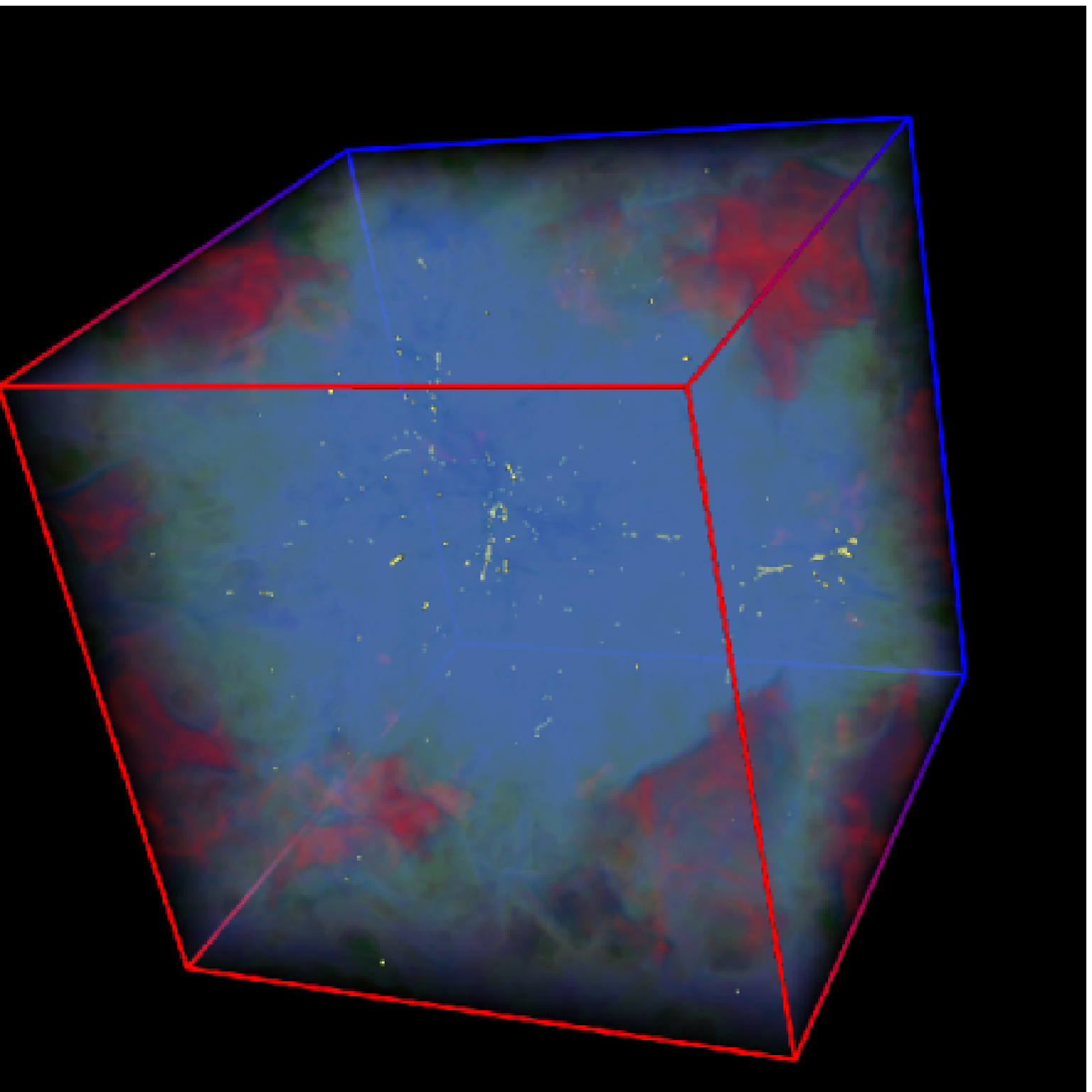}
\includegraphics[width=0.33\hsize]{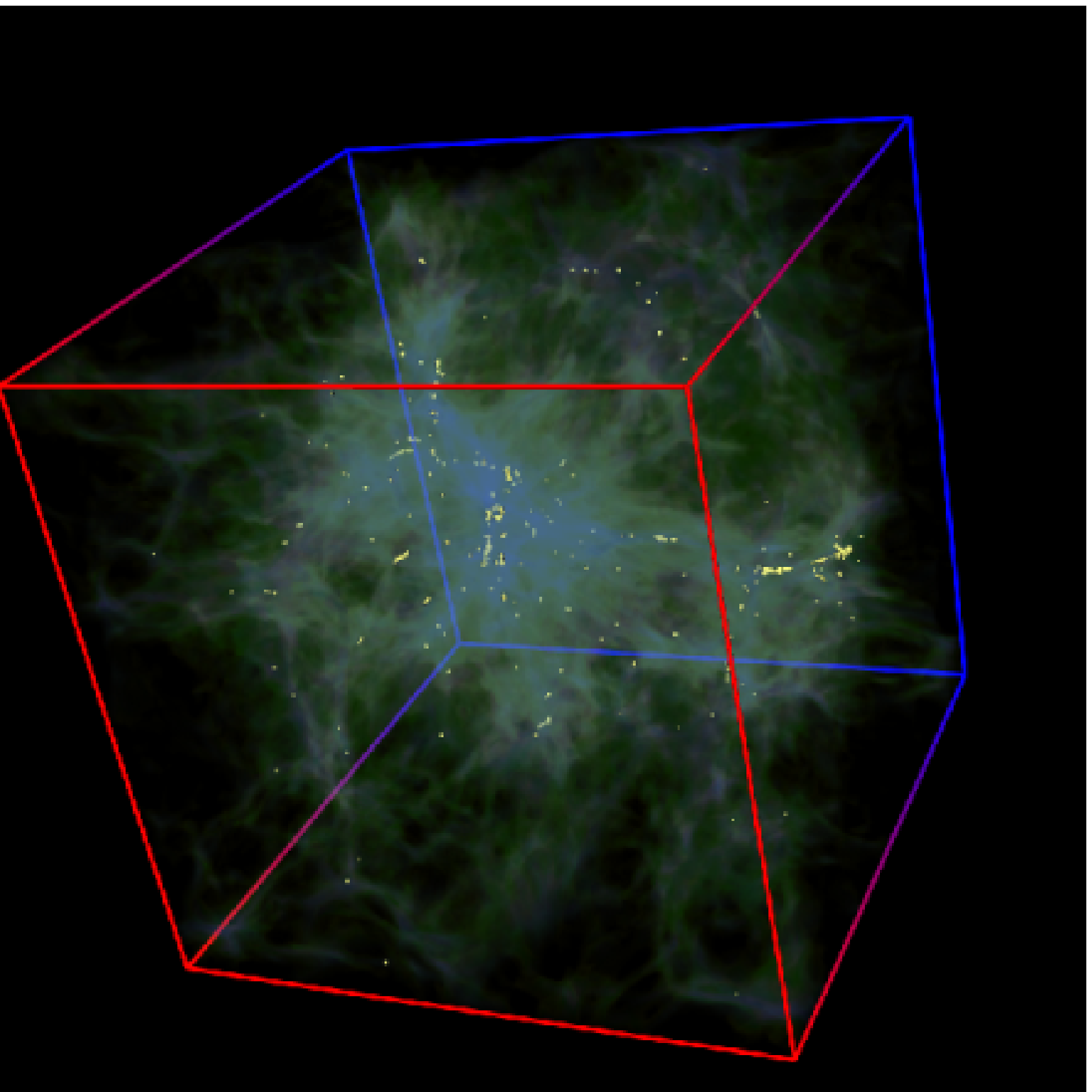}
\caption{\label{fig:sb} Visualization of the reionization process in
  small box simulations of \citet{ng:gf06}. Opaque brown material
  represents neutral hydrogen. Yellow points are galaxies. The glowing
  blue color shows recently ionized gas. The three panels show
  three different moments: $z=8.1$ (before the overlap of ionized
  bubbles), $z=6.3$ (beginning of overlap), and $z=5.5$ (after the
  overlap).} 
\end{figure*}

The primary challenge for modern simulations is to address the open question of how the distribution and properties of sources and sinks affect the reionization of the universe. As previously discussed, it is not yet possible to probe the full range of relevant scales with any single simulation. High-resolution simulations with small boxes can resolve radiation sinks such as Lyman limit systems, but the number of sources is not large enough to be representative of the actual galaxy distribution. On the contrary, large-box simulations have sufficient volume to be considered representative of the homogeneous and isotropic universe during the epoch of reionization, but generally have insufficient resolution to provide information on small-scale structure.

For clarity of discussion, we classify reionization simulations into two major categories: small-scale simulations that attempt to resolve known radiation sinks and large-scale simulations that attempt to account for the abundance of expected sources. In the first category, hydro + radiative transfer simulations directly model the evolution of the dark matter, cosmic gas, and radiation. These simulations resolve down to kpc scales or even smaller, allowing the modeling of two important physical processes. First, high-density, cooling gas can be found and turned into stars using simple, but physically plausible prescriptions \citep[e.g.][]{Katz1992, CenOstriker1992}. Second, the presence of high-density absorbing gas will act as Lyman limit systems.

Simulations falling into the second category share the common traits of having large volumes and high source counts. Hydro + radiative transfer simulations are predominantly not used, but instead hybrid approaches are taken. For example, high-resolution cosmological simulations are first run to model density fields and halo distributions. Radiative transfer calculations are then performed by post-processing the density fields using sources modeled from the halos. 
Various hybrid techniques have been developed, including those in which unresolved small-scale physics can be incorporated in approximate ways.

Studies of reionization necessarily overlap with many other directions of astrophysical and cosmological research. Constrained by the limited size of this review, we restricted our focus to computer simulations that model the whole reionization process and for which reionization is the prime subject of study. Thus, we do not discuss a large body of ground-breaking numerical work on modeling the formation of the very first stars in the universe \citep[see][for recent reviews]{BrommLarson2004, Norman2008}, and on simulating their effect on cosmic structure formation on small scales. Nor do we discuss an even larger body of numerical work on modeling the universe after reionization \citep[see][for a recent review]{cosmo:m07}. All of these simulations play an important role in developing our understanding of the high-redshift universe.

\subsection{Small-scale simulations that resolve sinks}
\label{sec:sbsims}

Small box simulations, that aim at resolving all sinks for ionizing
photons, have been historically the first attempt to 
model the process of reionization in 3D numerical simulations
\citep{ng:og96,ng:go97,sims:uns99,ng:g00a,Ciardi2001,sims:rnas02,Sokasian2003,sims:syah04}. 
As an example, the latest incarnation of these simulations performed by one
of us (NG) uses a box size of $8h^{-1}$ comoving Mpc and reach spatial
resolution 
of better than $700h^{-1}\dim{pc}$ in comoving units, or some
$100\dim{pc}$ at $8\la z\la 10$. These simulations reproduce well
the observed properties of Lyman limit systems at $z\sim4$ 
\citep{ng:kg07}. However, one can never be sure that some other type
of sinks exists during the reionization epoch; thus, the ``small-box''
simulations \textit{attempt} to resolve all sinks, but there is no
guarantee that they actually \textit{do} that.

An example of the small box simulations is shown in Fig.\
\ref{fig:sb}\footnote{More still images and several animations are
  available at
  \texttt{http://home.fnal.gov/\~{}gnedin/GALLERY/rei\_{}p.html}}. The
fact that the volume of these simulations is too small is immediately
apparent from the figure: most of reionization is done by just two
large ionized bubbles around two typical large galaxies (this is
especially apparent in the first panel). The rest of galaxies in the
simulation volume are small, dwarf galaxies; while they create a
larger number of smaller ionized bubbles around them, these bubbles do
not contribute significantly to the overall reionization process.

Obviously, two is not a statistically significant sample. In order to
partially compensate for the small box of these simulations, the
simulation volume is usually chosen not at random, but to be as close
to an average place in the universe as possible - such a choice
requires creating several hundreds of initial conditions and choosing
``the best one'' for the actual simulation. Because of such special
selection, these simulations can be used to reproduce some of the
observational data from SDSS quasars \citep{ng:gf06} that describe the
average properties of the universe; for example, the average fraction
of quasar light transmitted through the intergalactic medium, the mean
fraction of neutral hydrogen of the universe, etc. However, these
simulations fail when they are used beyond computing simple average
properties.

Another serious limitation of small box simulations is that
reionization in them always proceeds in the ``abundant sources''
scenario, simply because the simulation volume is smaller than the
mean free path due to Lyman limit systems. Ionized bubbles, then,
overlap before they reach sizes comparable to the mean free path and
never have a chance to stall, as in the ``rare sources'' scenario for
reionization. 

Numerical study of reionization began with small box simulations, but
by now they have largely completed their role. There is still one
application remaining for this type of simulations, though, and that
is their use in combination with very large volume simulations that
utilize the ``clumping factors'' approach, which we discuss next.

\subsubsection{Clumping factors approach}

\begin{figure}[t]
\begin{center}
\includegraphics[width=\hsize]{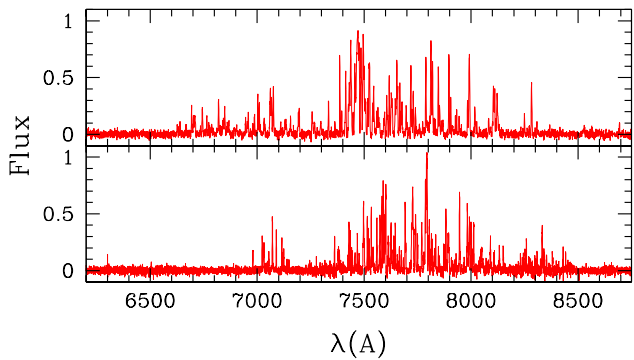}
\end{center}
\caption{\label{fig:fan2} A comparison of spectra of two $z=6.3$
  quasars: a real observed spectrum of the SDSS quasar J1030+0524 and
  a synthetic spectrum from a large-scale simulation with ``clumping
  factors'' approach. Guessing which is which we leave to the reader as
  a practical exercise.}
\end{figure}

Because the whole dynamic range, from Lyman limit systems to the
scale of cosmic horizon, is currently unachievable in a direct
numerical simulation, several approximate techniques have been
developed recently. While we do not overview all approximate
techniques here due to space limitations and our focus on simulations,
one of these techniques - the so-called ``clumping factors'' approach
-  still falls into a general simulation category. In this approach
the small-scales simulations that resolve Lyman limit systems are
combined with large-scale simulations that include a representative
volume of the universe (we discuss such simulations in the next
sub-section), or even a larger volume up to several Gpc on a side to
include the most rare quasars similar to ones observed by the SDSS
quasar survey.

The idea of clumping factors is simple. In a large-scale simulation of
a Gpc-sized volume the spatial resolution is limited to a few Mpc at
best, so a single resolution element (a mesh cell or a particle) in
such a simulation is a uniform region of a few Mpc on a
side. Certainly, the universe cannot be assumed to be uniform on this
scale; a ``clumping factor'' approach uses the small-scale simulations
to statistically describe the property of the universe on a few Mpc
scale, and then this statistical description is used in a large-scale
simulation as an approximation to the correct mathematical term in
the evolution equation. 

So far, the ``clumping factors'' approach has not been widely used
yet \citep{ng:kgh07,McQuinnHII2007}. Fig.\ \ref{fig:fan2} demonstrates the
ability of this approach to reproduce the observed spectra of high
redshift SDSS quasars, but only future work will demonstrate whether
the ability of ``clumping factors'' aproach to model extremely large
dynamic range (more than 1{,}000{,}000) justifies its approximate
nature and a substantial computational expense.

\subsection{Large-scale simulations that account for the abundance of sources}
\label{sec:lbsims}

\begin{figure*}[t]
\includegraphics[width=\hsize]{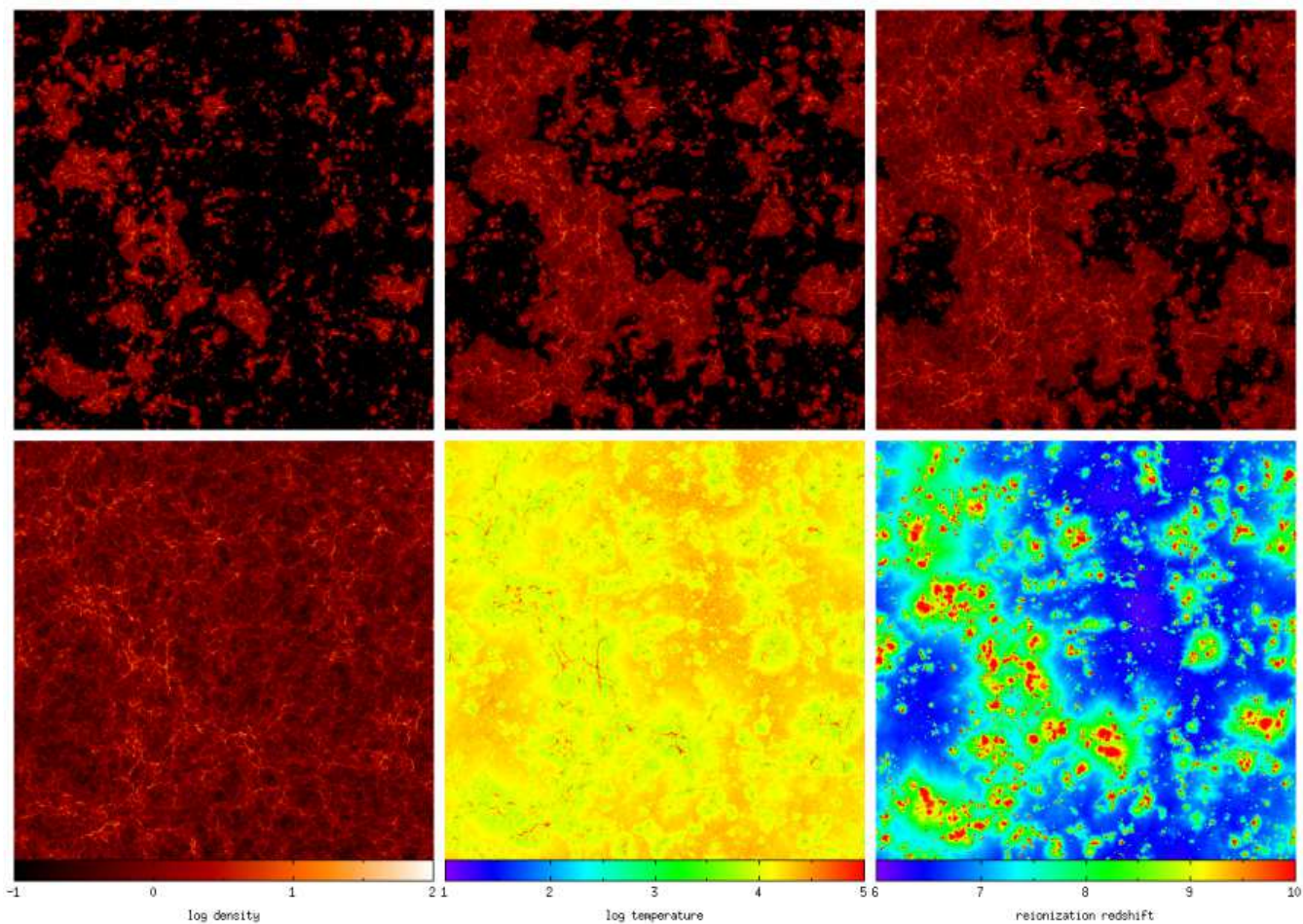}
\caption{\label{fig:trac08} Visualization of the reionization process in a hydro + radiative transfer simulation from \citet{TracCL2008}. The first four panels show the evolution of the ionized hydrogen density in a slice of size $(100\ \Mpch)^2$ when the simulation volume is 25, 50, 75, and 100 per cent ionized. The fifth panel shows the temperature at the end of reionization while the last panel shows the redshift at which gas elements get reionized. Higher-density regions tracing the large-scale structure are generally reionized earlier than lower-density regions far from sources. At the end of reionization, regions more recently photo-ionized and photo-heated are typically hotter because they have not yet had time to cool. Higher resolution images and movies are available at http://www.cfa.harvard.edu/$\sim$htrac/Reionization.html} 
\end{figure*}

Over this past decade, significant progress has been made towards understanding how the large-scale distribution of galaxies affects the reionization of the universe. Early simulations \citep[e.g.][]{Ciardi2003, Sokasian2003} were restricted to small box sizes of $10-20\ \Mpch$ and small numbers of sources because of the limited computing power then. In addition, the radiative transfer calculations were done in post-processing to simplify the computation. Despite the limitations, these simulations showed that the abundance and luminosities of galaxies clearly affected the reionization process and the development of ionized regions. The interesting results paved the way for continuing studies.

Since N-body simulations are less costly to run than hydro, they have been predominantly used in the last few years to generate density fields and halo distributions for hybrid modeling. The box size milestone of $100\ \Mpch$ for reionization simulations has now been reached \citep[e.g.][]{Iliev2006, Zahn2007, McQuinnLAE2007}, but only recently have both the box size and mass resolution requirements been met simultaneously \citep{Shin2008, IlievTG2008, TracCL2008}. However, the number of radiative transfer resolution elements used is considerably smaller because of the high computational cost.

Fig.\ \ref{fig:trac08} is a sample visualization from a modern large-scale simulation. \citet{TracCL2008} used a hybrid approach in which a high-resolution N-body simulation was first run to model sources and then hydro + radiative transfer simulations were run incorporating the sources. In the simulations, the radiative transfer of the ionizing photons proceeded such that large-scale, overdense regions near sources are generally photo-ionized and photo-heated earlier than large-scale, underdense regions far from sources.  The inhomogeneous process changed the thermal and ionization conditions, converting the cold and neutral gas into a warm and highly ionized one, typically with temperatures $\sim10^4\dim{K}$ and neutral fractions $\sim10^{-4}$.

\subsubsection{Ionized bubbles}

\begin{figure}[t]
\includegraphics[width=\hsize]{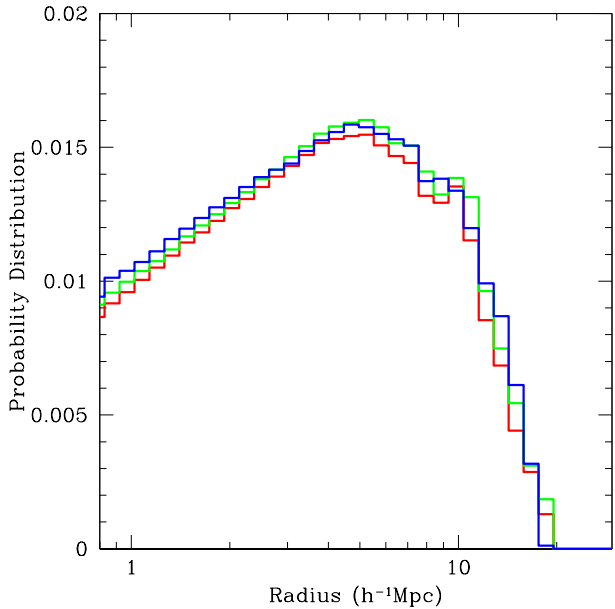}
\caption{\label{fig:bubble} Size distribution of ionized bubbles when reionization is half completed. The results, from two ray-tracing codes (McQuinn et al.~2007b in red; Trac \& Cen 2007 in green) and a semi-analytic code (Zahn et al.~2007 in blue), are in good agreement overall. They have similar characteristic peak sizes, but some differences are seen at small and large radii. Figure is courtesy of O.~Zahn.} 
\end{figure}

The large-scale distribution and properties of ionized bubbles have been studied in more detail in the last few years. Recent simulations \citep[e.g.][]{Iliev2006, Zahn2007, McQuinnHII2007, Lee2008, Shin2008, CroftAltay2008} support the basic picture for the development of ionized bubbles described in \S \ref{sec:physics}. Ionized bubbles originate within halos hosting sources and expand outwards, such that higher-density regions near sources are ionized earlier than lower-density regions far from sources. Bubbles will merge with one another until they overlap and fill all of space.

\citet{McQuinnHII2007} and \citet{CroftAltay2008} have simulated a range of models for sources and sinks and are in agreement that sources more strongly influence the development of ionized bubbles. The properties of sources can affect the morphology in several major ways. Ionized bubbles are generally aspherical, as seen in Fig.\ \ref{fig:trac08}, but rarer sources tend to generate larger and more spherical bubbles. Being more strongly clustered, rare sources have many other smaller sources nearby that contribute photons and help ionize a larger volume. On the other hand, lower luminosity sources can significantly add to the abundance of small bubbles, if they are not already embedded within larger bubbles. The lower limit on luminosity will depend on the efficiency of star formation in smaller mass halos.

While not the dominant factor, the distribution of sinks can still have strong effects on reionization. The abundance of sinks, from Lyman limit systems and minihalos, restricts the mean free path of ionizing photons. Once the mean free path is shorter than the bubble size, the ionized regions stop growing, thus delaying the reionization process. In the presence of stronger sinks, the distribution of bubble sizes shifts towards smaller bubbles.

For a partially ionized universe, the probability distribution or histogram of bubble size generally increases towards smaller sizes. There are relatively more small bubbles than large bubbles, reflective of there being more low-mass halos than high-mass halos. This is the nature of the hierarchical universe in which larger and rarer objects are assembled from smaller and more abundant objects. Note that there is not a one-to-one correspondence between bubbles and sources since ionized regions can have one or more galaxies embedded within. The bubble size distribution can also be calculated by weighting each bubble by its volume rather than a simple count. In this version, the distribution has a characteristic peak size that shifts toward larger scales as reionization progresses and ionized bubbles grow \citep{FurlanettoZH2004}.

Fig.\ \ref{fig:bubble} shows a sample result on the bubble size distribution from a code comparison project. Two ray-tracing codes \citep{McQuinnHII2007, TracCen2007} and a semi-analytic code \citep{Zahn2007} have calculated the reionization process, starting from identical initial conditions of a $100\ \Mpch$ box. The results shown are taken at the same redshift, when each simulation has reionized half the volume. The three calculations are in good agreement overall, including having the same characteristic peak size, but small differences are present due to the different treatments for radiative transfer.

In the last few years, semi-analytic models for reionization \citep[e.g.][]{BarkanaLoeb2004, FurlanettoZH2004, Zahn2007, MesingerFurlanetto2007, ChoudhuryHR2009} are predominantly based on a technique called the `excursion set formalism' \citep{Bond1991}. In these schemes, the source distributions and ionization fields are approximately derived from the density fields. Basically, sources are first located within the density field by asking whether there is enough surrounding mass at a given point in space. Ionized bubbles are then found by asking whether a given region has enough sources within it to ionize the contained mass. Since the propagation of photons is not directly followed, the algorithms tend to be much faster and use less memory than radiative transfer algorithms. Semi-analytic models have been demonstrated to be in good agreement with radiative transfer simulations, but there still are differences to be worked out. They are very useful for studying reionization, especially when a large number of models needs to be run.

\subsubsection{21 cm radiation}
\label{sec:21cm}

\begin{figure*}[t]
\includegraphics[width=\hsize]{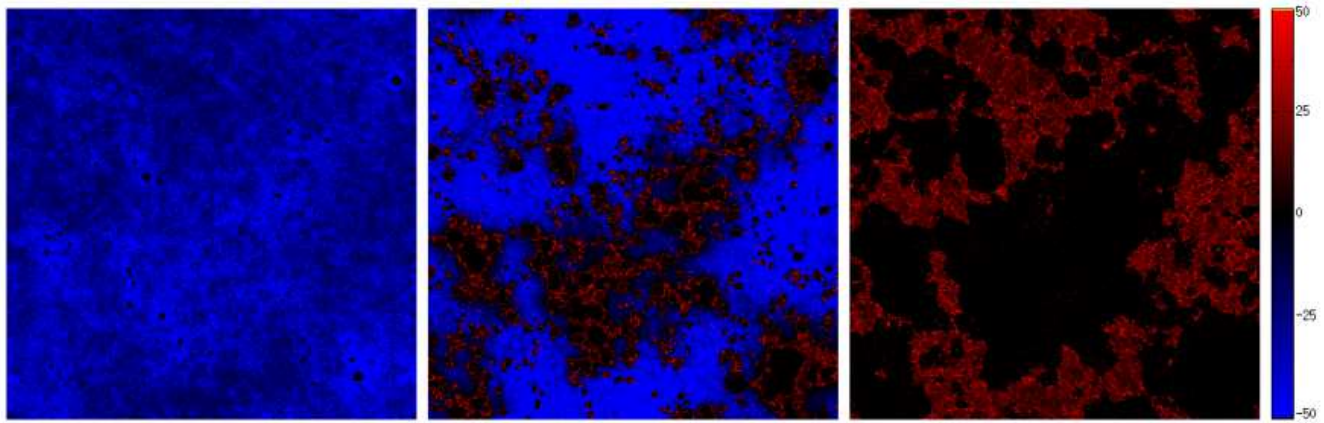}
\caption{\label{fig:santos08} Maps of the 21\dim{cm} brightness temperature (mK) showing the transition from absorption (blue) to emission (red) in a slice of size $(100\ \Mpch)^2$, taken from the simulation of \citet{Santos2008}.
The first panel shows the absorption signal from a very early stage of reionization when the spin temperature was less than the CMB temperature. In the second panel, neutral regions, surrounding ionized bubbles, that have been heated by high-energy X-ray photons have switched from absorption to emission. The last panel shows the emission signal when the simulation volume is half ionized, but the spin temperature exceeds the CMB temperature everywhere. Figure is courtesy of Alexandre Amblard and Mario Santos.} 
\end{figure*}

Neutral hydrogen from the epoch of reionization can be detected through high-sensitivity radio observations. In a hydrogen atom, the proton and electron both have spins and the alignment of the spins can flip between being parallel and antiparallel. This spin flip, sometimes referred to as a hyperfine transition, has an energy level difference corresponding to a rest wavelength of 21\dim{cm}. During reionization, this signal is seen in emission or absorption relative to the CMB and the brightness of the signal can be used to probe the characteristics of neutral regions and the delineation with ionized regions. Studying the epoch of reionization and the high redshift universe with 21\dim{cm} radiation is a frontier topic in cosmology. We refer the reader to \citet{FurlanettoOB2006} and \citet{Loeb2008} for recent reviews.

Recent simulations \citep[e.g.][]{Mellema21cm2006, Lidz2007, Santos2008, Baek2009} have studied the contrast between neutral and ionized regions by calculating the expected 21\dim{cm} brightness, usually expressed as a ``brightness temperature'', the quantity widely used in Radio Astronomy. The brightness temperature depends on the neutral hydrogen density and the gas spin temperature, which characterizes the fraction of parallel to antiparallel spin states. The signal is seen in emission when the spin temperature exceeds the CMB temperature and in absorption if the opposite is true.

The evolution of the gas spin temperature depends on three mechanisms: collisions within the gas, scattering with the CMB, and scattering with Lyman alpha photons. The latter dependence is commonly known as the Wouthysen-Field effect \citep{Wouthuysen1952, Field1959}. More recent simulations have made the effort to calculate the evolution of the spin temperature. \citet{Santos2008} calculated the rise of the spin temperature by including heating of neutral gas by high-energy X-rays and pumping of spin states by Lyman alpha photons, as sources turn on. \citet{Baek2009} performed a more exact calculation of the radiative transfer of Lyman alpha photons, taking into account the resonant scattering with neutral hydrogen. These inclusions are important in earlier stages of reionization when sources are few. 

Fig.\ \ref{fig:santos08} shows the transitioning of the 21\dim{cm} brightness temperature from absorption to emission in the simulation of \citet{Santos2008}. Initially, the signal is seen in absorption because the neutral hydrogen gas is cold and the spin temperature is less than the CMB temperature. As sources turn on and produce radiation, the signal vanishes in the ionized regions. The X-rays and Lyman alpha photons will travel beyond the ionized regions and raise the spin temperature of surrounding neutral regions, resulting in the transition from absorption to emission. As reionization progesses, the emission regions will shrink in size.

The primary statistic for studying the 21\dim{cm} signal has been the power spectrum of the brightness temperature field. The power spectrum quantifies how fluctuations in the brightness field are correlated with one another. The power spectrum of a field $\delta(\bx)$ is normally obtained by first Fourier transforming the field and then calculating the average power $\langle |\delta(\bk)|^2 \rangle$ for modes with wavenumber $k=|\bk|$. It is also the Fourier transform of the two-point correlation function, which was discussed in \S \ref{sec:physics}, and thus, a measure of spatial correlations. While the 21\dim{cm} signal comes from the neutral regions, the relative fluctuations in the brightness field are commonly discussed in the context of ionized bubbles.

The 21\dim{cm} power spectrum has some interesting features. On scales large compared to the characteristic bubble size, the 21\dim{cm} power spectrum is approximately proportional to the power spectrum of the matter. This correlation is due to the fact that ionized bubbles approximately trace the large-scale structure, as seen in Fig.\ \ref{fig:trac08}. The scaling or bias factor comes from the fact that bubbles can cluster differently than the matter. The power spectrum also has a characteristic peak scale corresponding to the characteristic bubble size. As reionization progresses and the bubbles merge, the peak shifts toward larger scales and the power spectrum changes shape. Several groups of radio astronomers are actively developing instruments to measure this signal early in the next decade. These observations will provide a wealth of information on the reionization process and test our theoretical concepts in great detail \citep[e.g][]{FurlanettoOB2006, Lidz2008, PritchardLoeb2008, Barkana2008}.

\section{Conclusions}

Computer simulations of reionization have progressively achieved better physical realism and dynamic range over the past decade. The constant improvement of numerical algorithms combined with the rapid advancement in supercomputing resources have fueled the growth. With N-body and hydro simulations, we can now robustly evolve the dark matter and cosmic gas to form the large-scale structure of the universe. In conjunction, radiative transfer algorithms, based on moments, Monte Carlo, and ray-tracing methods, have enabled us to accurately solve the propagation of ionizing photons. While various implementations exist, good algorithms possess the important features of conserving photons and scaling linearly with the number of resolution elements.

Armed with modern tools, we have made significant strides in understanding how the universe was reionized and how sources and sinks influenced the process. The development of ionized bubbles, first around sources, followed by expansion to merge with other bubbles, and finally overlapping all of space, is now a quantifiable theory. Statistical tools have been developed to characterize the morphology and clustering of bubbles, and to discriminate between the possible scenarios for reionization.

One of the major challenges remaining is to develop better physical models for sources and sinks that can be tested against high-redshift observations. Ideally, our ignorance of complex astrophysical processes can be represented by a small number of unknown, but constrainable, parameters. The task of a computational cosmologist then becomes to ensure the high fidelity of the numerical solution given these parameters, so that modeling the process of reionization reduces to searching the parameter space. Since reionization is a complex process with no general analytical solution, the check on simulations requires having numerical codes converge with one another.

We will have better observational constraints on reionization from ongoing surveys of high-redshift quasars and young, star-forming galaxies. The latter is especially important for learning about sources. Upcoming radio observations of the 21\dim{cm} radiation from neutral hydrogen have the potential to be the best probe of the epoch of reionization. In addition, higher-resolution experiments at microwave wavelengths will be able to measure the fluctuations in the CMB temperature caused by scattering with electrons produced by the reionization process. A wealth of information is expected over the next decade. Continuing progress on the theoretical front is therefore necessary to compliment the exciting promise of future observations.

\acknowledgments

We thank Alexandre Amblard, Xiaohui Fan, Ilian Iliev, Mario Santos, Volker Springel, and Oliver Zahn for kind permission to use their figures in this review. H.T.\ is supported by an Institute for Theory and Computation Fellowship. N.G.\ is supported in part by the DOE at Fermilab.

\bibliography{ng-bibs/cosmo,ng-bibs/igm,ng-bibs/self,ng-bibs/qlf,ng-bibs/atom,ng-bibs/hizgal,ng-bibs/sims,trac}

\end{document}